\documentclass[conference]{IEEEtran}
\IEEEoverridecommandlockouts
\usepackage{cite}
\usepackage{amsmath,amssymb,amsfonts}
\usepackage{algorithmic}
\usepackage{graphicx}
\usepackage{textcomp}
\usepackage{xcolor}
\usepackage{listings}
\usepackage{url}
\usepackage{hyperref}
\usepackage{todonotes}
\usepackage{xcolor}
\def\BibTeX{{\rm B\kern-.05em{\sc i\kern-.025em b}\kern-.08em
    T\kern-.1667em\lower.7ex\hbox{E}\kern-.125emX}}
\begin{document}

\title{Trajectools Demo: Towards No-Code Solutions for Movement Data Analytics
\thanks{This work is mainly funded by the EU's Horizon Europe research and innovation program under Grant No. 101093051 EMERALDS.}
}

\author{\IEEEauthorblockN{Anita Graser}
\IEEEauthorblockA{\textit{AIT Austrian Institute of Technology}\\
Vienna, Austria \\
ORCID: 0000-0001-5361-2885}
\and
\IEEEauthorblockN{Melitta Dragaschnig}
\IEEEauthorblockA{\textit{AIT Austrian Institute of Technology} \\
Vienna, Austria \\
ORCID: 0000-0001-5100-2717}
}

\maketitle

\begin{abstract}
This demo paper presents the conceptual foundations and the first steps towards implementation of a novel no-code solution for movement data analytics based on the open-source Python library MovingPandas and the open-source geographic information system QGIS. 
The resulting Trajectools plugin is available open-source at \url{https://github.com/movingpandas/qgis-processing-trajectory}.
\end{abstract}

\begin{IEEEkeywords}
mobility data science, geographic information science, trajectories, visual analytics
\end{IEEEkeywords}

\section{Introduction}

In recent years, multiple research groups have been working on scientific software libraries for movement data analytics in Python~\cite{graser2019,pappalardo2022,martin2023}, R~\cite{joo2020}, and other languages\footnote{\url{https://github.com/anitagraser/movement-analysis-tools}}.
However, these libraries are still rather fragmented (no single library covers all use cases) and their usage requires significant programming skills. This makes them difficult to learn and thus limits their use as well as the transfer of research advances to other disciplines, industry, and wider public. 
Furthermore, the visualization capabilities of these libraries are limited~\cite{graser2022} which presents a further challenge since the mobility and ``transportation domain is characterized by both complex data and complex problems, which calls for visual analytics approaches''~\cite{andrienko2017}.

Geographic information systems (GIS) provide strong capabilities for analyzing and visualizing spatial data and are therefore often used to work with movement data and related geographic context data in the mobility and transport domain. 
GIS-based no-code solutions such as the ArcGIS Model Builder\footnote{\url{https://pro.arcgis.com/en/pro-app/latest/help/analysis/geoprocessing/modelbuilder/what-is-modelbuilder-.htm}}, the CARTO Workflow design tool\footnote{\url{https://carto.com/workflows}}, or the QGIS Model Designer\footnote{\url{https://docs.qgis.org/3.28/en/docs/user_manual/processing/modeler.html}} are popular ways to create reproducible workflows for spatial data analysis \cite{graser2014,cosentino2016,raimondi2023}. 
However, so far, these no-code solutions do not extend to spatio-temporal movement data.
Other widely accessible visual analytics systems' capabilities are also very limited with respect to spatio-temporal data\cite{andrienko2020}.
To achieve a better understanding of environmental, economic, and societal processes, researcher need to ``not only strive for advancing the research but also take the responsibility for transferring the operational  knowledge to practitioners and casual analysts''\cite{andrienko2020}.

The integration of movement data analysis libraries into GIS-based no-code solutions is hindered by a mismatch between movement data models for trajectories and classic GIS data models following the OGC Simple Features (SF) standard\footnote{\url{https://www.opengeospatial.org/standards/sfa}}. 
SF defines point, line, and polygon features that combine geometry and arbitrary attributes but the temporal dimension is not standardized. 
Mobility data specific standards, such as OGC Moving Features\footnote{\url{https://www.ogc.org/standards/movingfeatures}}, have not seen wide adoption yet~\cite{graser2020}. 

The following sections present the concepts behind Trajectools, enabling the integration of  movement data analysis libraries into GIS-based no-code solutions and demonstrate its use in a real-world use case.

\section{Trajectools Concept \& Implementation}

\begin{figure}[htbp]
\centerline{\includegraphics[width=1\linewidth]{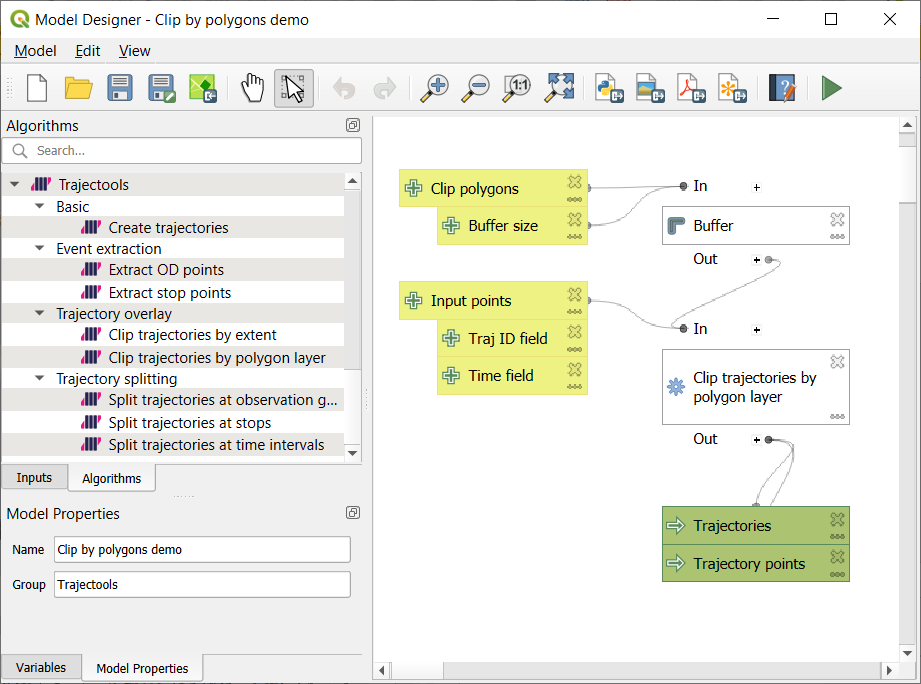}}
\caption{QGIS Processing Model Designer with the list of Trajectools algorithms on the left and an example model combining Trajectools \textit{Clip trajectories by polygon layer} algorithm with QGIS' default buffer algorithm}
\label{fig-toolbox}
\end{figure}

Trajectools is a plugin that adds trajectory algorithms to the QGIS Processing Toolbox~\cite{graser2015} by wrapping the functionality of the MovingPandas library~\cite{graser2019}. 
The latest version of Movingpandas (v0.17) implements functions for trajectory interpolation, re-sampling, splitting, extraction of events, and aggregation. Trajectools currently wraps a subset of these functions (as shown in Figure~\ref{fig-toolbox}) and more are being added as the development progresses. 
This implementation profits from the open-source nature of both tools and the fact that QGIS is easily extendable with plugins written in Python.

Raw trajectory data that serves as input for Trajectools is usually provided as points with timestamps and trajectory IDs in formats such as CSV, GPX, or in database formats that can be read as a GIS point layer (e.g. PostGIS or MobilityDB~\cite{graser2018,zimanyi2020}). 
A single dataset or resulting GIS point layer may describe one or more trajectories. 
Trajectory algorithms can take these raw trajectory points, process them and output either processed trajectories or derived information, such as significant trajectory points (including origin and destination (OD) points, turn points, and stop points) or aggregated information (such as OD flows and flows between turning points).

To overcome the lack of standardized movement data handling, Trajectools follows a pragmatic approach relying on the SF-based vector layer implementations commonly provided by GIS. The first step is to define the interfaces, that is, how trajectory data is to be passed between the modular building blocks of the model workflows. It is essential to ensure that the names of trajectory id and timestamp attributes are not altered when creating the output since changing attribute names would hinder the chaining of tools.  Therefore, the base class \textit{TrajectoriesAlgorithm} implements the common input parameters (input point layer, trajectory ID field, timestamp field, speed units). Derived classes, such as \textit{TrajectoryManipulationAlgorithm} and \textit{ClipTrajectoriesByPolygonLayer} extend the base classes' inputs and outputs. 
Timestamps formats and Coordinate reference system (CRS) information are inferred automatically wherever possible. 


\textbf{Trajectory manipulation algorithms} take trajectory point data as input and create trajectory outputs in the form of points and linestrings. These different output types facilitate different follow-up analysis and visualization steps. They allow the user to chain multiple algorithms using the point output of one tool as input for the next. 
The currently implemented trajectory manipulation algorithms include:

\begin{itemize}
    \item \textbf{Basic} / \textit{Create trajectories}  creates a trajectory point layer with speed and direction information as well as a trajectory line layer with start and end time, duration, length, and average speed. 
    \item \textbf{Trajectory Overlay} algorithms operate on two input layers to produce an output layer combining information from both inputs, including \textit{Clip trajectories by polygon layer} and \textit{Clip trajectories by extent} where the extent can be specified manually, set from the current map extent, or derived from a map layer, layout, or bookmark. 
    \item \textbf{Trajectory Splitting} algorithms split input trajectories into subtrajectories based on different spatial or spatio-temporal criteria, including \textit{Split trajectories at observation gaps}, \textit{Split trajectories at stops}, and \textit{Split trajectories at time intervals} (year, month, day, or hour).
\end{itemize}

The trajectory points output attribute table is automatically populated with speed and direction values using mpd.TrajectoryCollection.add\_speed and add\_direction functions\footnote{\url{https://movingpandas.readthedocs.io/en/main/trajectorycollection.html}}. 
The speed units are customizable using MovingPandas' unit support\footnote{\url{https://movingpandas.org/units}}.
The trajectory output attribute table is similarly populated with movement information, including trajectory start and end time, duration, length, and average speed.

\begin{figure}[htbp]
\centerline{\includegraphics[width=1\linewidth]{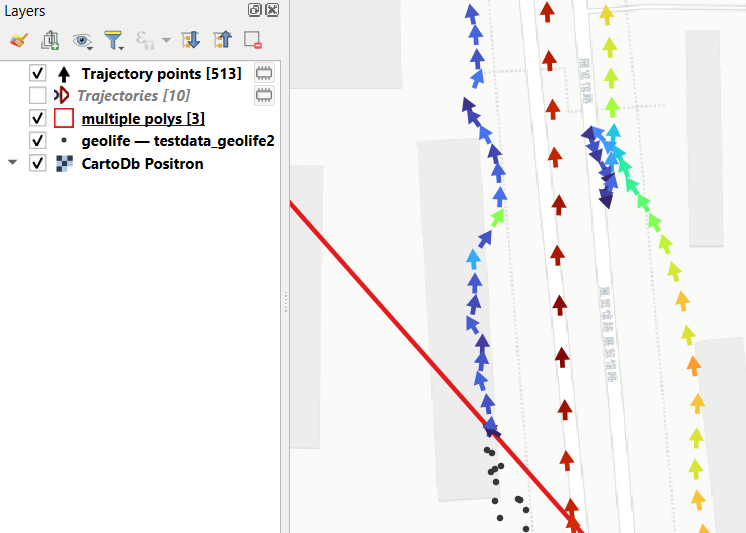}}
\caption{Details of the points output layer of the \textit{Clip trajectories by polygon layer} algorithm. The  blue-yellow-red ``Turbo'' gradient visualizes the computed speeds from slow (blue) to high (red).}
\label{fig-clip-results-pts}
\end{figure}

\begin{figure}[htbp]
\centerline{\includegraphics[width=1\linewidth]{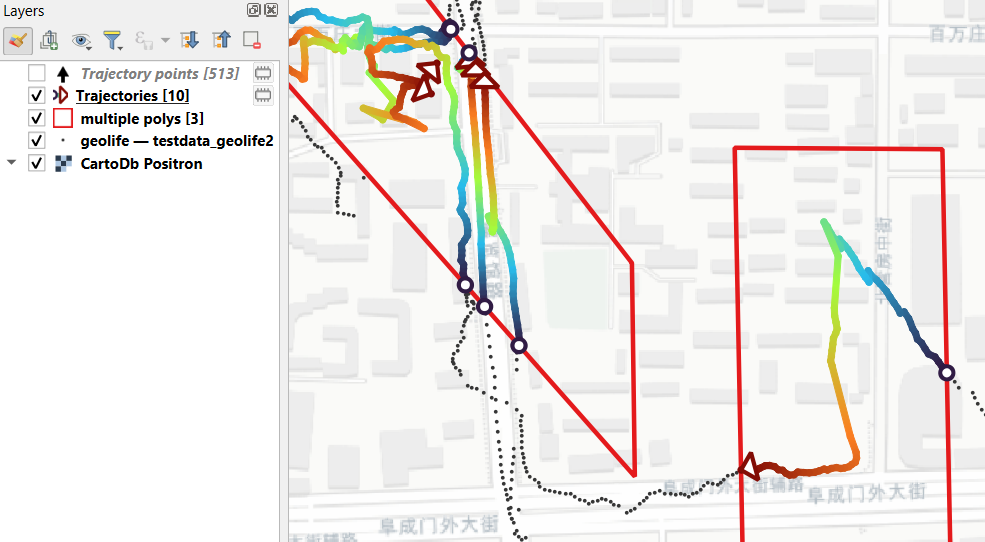}}
\caption{Details of the trajectories output layer of the \textit{Clip trajectories by polygon layer} algorithm. The  blue-yellow-red ``Turbo'' gradient visualizes the trajectory direction from start (blue) to end (red).}
\label{fig-clip-results-traj}
\end{figure}

\begin{figure*}[htbp]
\centerline{\includegraphics[width=0.85\textwidth]{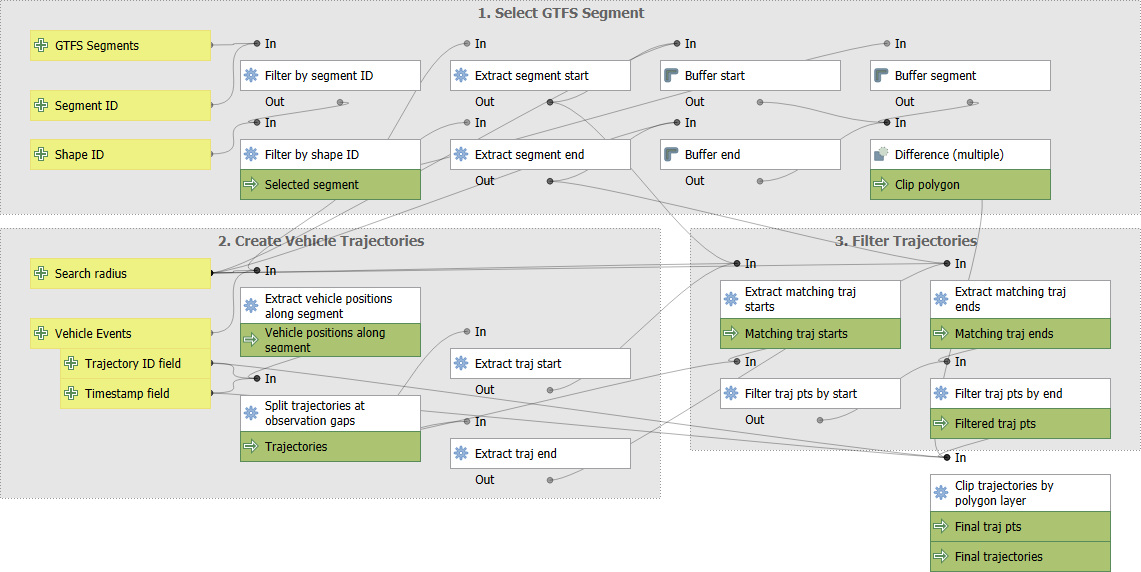}}
\caption{Processing model for the public transport travel time analysis example.}
\label{fig-riga-model}
\end{figure*}

To enable efficient visual movement data analytics, the output layers are automatically styled to visualize movement information. The point style visualizes computed speed and direction (see Figure~\ref{fig-clip-results-pts}) and the trajectory style visualizes movement direction (see Figure~\ref{fig-clip-results-traj}). For the point style, it is necessary to make a trade-off between two approaches: by keeping the minimum and maximum speed values for the color gradient fixed, it is straightforward to compare outputs from different algorithm runs. However, the hard-coded minimum and maximum speed values (currently 0 and 50km/h) may not be suitable for all applications. Alternatively, the minimum and maximum values could be determined automatically based on the output layer's speed values at the cost of losing immediate comparability of outputs. Since it is easy to alter the speed limits (if necessary), the first approach with hard-coded limits has been implemented.

\textbf{Event extraction algorithms}, on the other hand, create non-trajectory outputs, such as significant points, flows between significant points, or home ranges (the area in which an animal lives and moves on a periodic basis) which can be represented as points, lines or polygons, respectively. These outputs can be used as inputs for other algorithms but cannot be used as trajectory inputs for trajectory algorithms.  The currently implemented extraction algorithms include:  \textit{Extract OD points} and  \textit{Extract stop points}.




\section{Demonstration}

Trajectools is available from the QGIS Plugin repository\footnote{\url{https://plugins.qgis.org/plugins/processing_trajectory/}} for anyone to use with their own data or with the included sample data. 
In our demo, we present the use of Trajectools in a real-world use case from the EMERALDS project: the analysis of travel times of public transport (PT) vehicles. In this use case, PT vehicles are tracked via GPS. However, the logged vehicle tracks cannot be directly linked with PT route segments from the PT schedule provided as General Transit Feed (GTFS) due to a lack of common keys. Therefore, the implemented model performs spatial join and filter operations to identify the PT trajectory segments that followed a specific GTFS route segment (i.e. the route between two consecutive PT stops). 
The model presented in Figure~\ref{fig-riga-model} consists of four main blocks: 
Block~1 extracts the user-selected GTFS segment and processes it to provide the segment's start and end locations as well a polygon mask that is used later in the model to clip the trajectories. 
Block~2  processes the PT vehicle tracks and creates candidate trajectories of all vehicle locations within a user-specified search radius of the selected GTFS segment. 
Block~3  filters down the candidate trajectories from step~2 by ensuring that they start and end close to the GTFS segment's start and end locations, respectively. (This ensures the correct movement direction of the trajectories.)   
Finally, we clip the filtered trajectories at the start and end locations to ensure that waiting times at PT stops don't affect the computed segment travel speeds.

\begin{figure}[htbp]
\centerline{\includegraphics[width=0.7\linewidth]{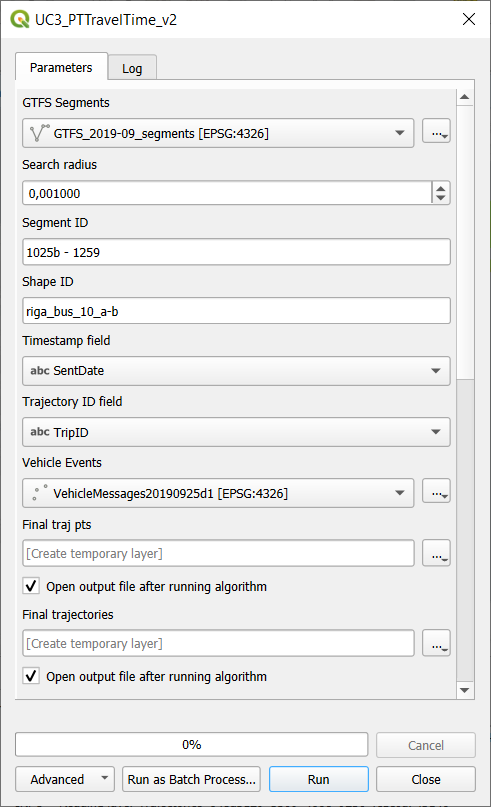}}
\caption{Automatically generated Processing model GUI for the public transport travel time analysis example.}
\label{fig-riga-model-gui}
\end{figure}

Yellow boxes indicate model inputs, white boxes are the algorithms, and green boxes are the model outputs. This model creates many outputs representing intermediate results which help understand and debug the analysis process. When running the model, these outputs can be deactivated by unchecking the ``Open output file after running algorithm'' checkboxes in the GUI, shown in Figure~\ref{fig-riga-model-gui}.

The Trajectools algorithms applied in this model are: \textit{Split trajectories at observation gaps} and \textit{Clip trajectories by polygon layer}. All other algorithms used in this model are default algorithms provided by the Processing Toolbox. This example thus illustrates the power of building on top of an existing GIS-based framework, instead of attempting to reinvent the wheel, since spatial analysis steps and attribute-based joins can be performed with existing algorithms and interleaved with new Trajectools algorithms.

\begin{figure}[htbp]
\centerline{\includegraphics[width=1\linewidth]{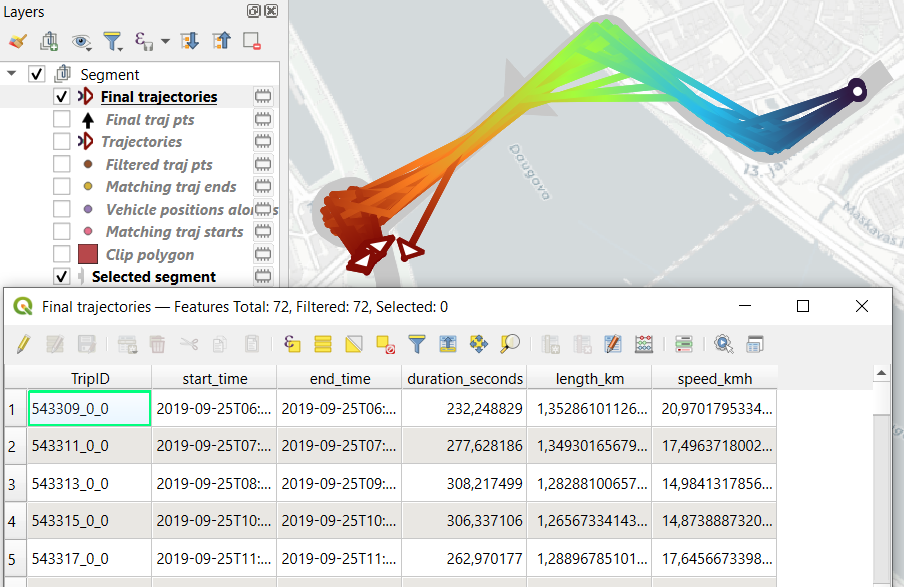}}
\caption{Output of the public transport travel time analysis model for a select GTFS segment in Riga.  }
\label{fig-riga-results}
\end{figure}

Figure~\ref{fig-riga-results} shows the resulting trajectories on the map as well as the resulting attribute table with the computed average segment speed values. One caveat is that the computed trajectory speeds are impacted by the rather sparse reporting interval (which can be seen on the map where the trajectories ``cut corners''). Therefore, the speeds should be corrected by using the actual segment length before comparing the observed speeds with the scheduled speeds provided by the GTFS feed (between 21-24~km/h depending on the time of day).

\section{Outlook}

The conceptual work for no-code solutions for mobility data analysis presented in this paper and the Trajectools plugin implemented in QGIS Processing can serve as a prototype for how to add mobility support in other GIS-based no-code environments. 
More integration work is ongoing to achieve feature parity between Trajectools and MovingPandas by adding trajectory cleaning, resampling, smoothing, and aggregation. Simultaneously, feedback from Trajectools users may lead to the development of new algorithms for MovingPandas.

Our vision for future work includes the integration of algorithms from related libraries, such as scikit-mobility~\cite{pappalardo2022} and Trackintel~\cite{martin2023} into Trajectools to allow for a seamless usage of these algorithms within the no-code environment. We expect that this integration will be helped by the fact that these libraries also build on the (Geo)DataFrame datatype which should make it relatively straightforward to exchange data between libraries.

\section*{Acknowledgment}

The examples shown in Figure~\ref{fig-clip-results-pts}-\ref{fig-clip-results-traj} use data extracted from the public Geolife dataset~\cite{zheng2010}.
The Riga example data was kindly provided by Riga's public transport company Rīgas Satiksme.

\vspace{12pt}

\end{document}